\documentclass[a4paper,fleqn,usenatbib]{mnras}

\usepackage{newtxtext,newtxmath}

\usepackage[T1]{fontenc}
\usepackage{ae,aecompl}

\usepackage{graphicx}	
\usepackage{amsmath}	
\usepackage{amssymb}	
\usepackage{color,soul}
\soulregister\cite7

\title[Formation of stellar clusters]{Formation of stellar clusters}

\author[R. Smilgys \& I. A. Bonnell]{
Romas Smilgys,$^{1}$\thanks{E-mail: rs202@st-andrews.ac.uk}
Ian A. Bonnell$^{1}$
\\
$^{1}$Scottish Universities Physics Alliance (SUPA), School of Physics and Astronomy, 
University of St. Andrews, North Haugh, St. Andrews, Fife KY16 9SS, UK
}

\date{Accepted XXX. Received YYY; in original form ZZZ}

\pubyear{2017}

\begin{document}
\label{firstpage}
\pagerange{\pageref{firstpage}--\pageref{lastpage}}
\maketitle

\begin{abstract}

We investigate the triggering of star formation and the formation of stellar clusters in molecular clouds that form as the ISM passes through spiral shocks. The spiral shock compresses gas into $\sim$100 pc long main star formation ridge, where clusters forming every 5-10 pc along the merger ridge. We use a gravitational potential based cluster finding algorithm, which extracts individual clusters, calculates their physical properties and traces cluster evolution over multiple time steps. Final cluster masses at the end of simulation range between 1000 and 30000 M$_{\odot}$ with their characteristic half-mass radii between 0.1 pc and 2 pc. These clusters form by gathering material from 10-20 pc size scales. Clusters also show a mass - specific angular momentum relation, where more massive clusters have larger specific angular momentum due to the larger size scales, and hence angular momentum from which they gather their mass. The evolution shows that more massive clusters experiences hierarchical merging process, which increases stellar age spreads up to 2-3 Myr. Less massive clusters appear to grow by gathering nearby recently formed sinks, while more massive clusters with their large global gravitational potentials are increasing their mass growth from gas accretion.

\end{abstract}




\section{Introduction}

Star formation is one of the most important processes in galactic evolution, transforming gas into stars and providing the visible output as well as the chemical and energetic feedback into the galaxy. Current understanding is that at least half of all stars, and potentially all massive stars, form in stellar clusters (\citealt{2003ARA&A..41...57L, 2004A&A...425..937D, 2007ARA&A..45..481Z, 2010MNRAS.409L..54B, 2015ApJ...805...92O, 2016AJ....151....5M, 2017ApJ...834...94S}). Understanding how stellar clusters form is on ongoing challenge with many observational studies trying to determine the exact initial conditions (\citealt{2010ApJ...723L...7K, 2017MNRAS.468.3694C, 2017A&A...601A..60C}). Models for cluster formation show that a turbulent molecular clump that is gravitationally unstable fragments hierarchically into small clusters which eventually merge to form larger stellar systems (\citealt{2003MNRAS.343..413B}, \citealt{2003MNRAS.339..577B}, \citealt{2007arXiv0712.0828K}, \citealt{2008ApJ...686.1174O}, \citealt{2009MNRAS.400.1775S}). Observations of star formation in infrared dark clouds such as \cite{2013A&A...555A.112P} shows the fragmentation of the filamentary structure (\citealt{2014prpl.conf...27A}, \citealt{2017arXiv170307029H}) feeding into the formation of a stellar cluster.

Numerical simulations have provided significant insight into the dynamical nature of star formation (\citealt{2003MNRAS.339..577B}; \citealt{2003MNRAS.343..413B}; \citealt{2011MNRAS.410.2339B}; \citealt{2016MNRAS.462.4171B}), showing the importance of turbulence, collapse, fragmentation, interactions and accretion. Although useful in highlighting the physical processes, these numerical simulations suffer from their overly-idealised initial conditions. Our (\citealt{2011MNRAS.410.2339B}) study showed that a small variation in gravitational boundedness along a cloud results in significantly different stellar populations, clusterings, star formation rates, efficiencies, and stellar IMFs. Initial conditions have a major impact on all star formation properties, and hence using self-consistent initial conditions is crucial to develop realistic models.

Recent simulation work, like \cite{2003MNRAS.339..577B}, \cite{2003MNRAS.343..413B}, \cite{2011MNRAS.410.2339B}, \cite{2011ApJ...740...74K} gives results for individual molecular clouds and forming clusters. Observations such as those of \cite{2015ApJ...802..125R} support the concept that clusters grow hierarchically. However, observations are limited in terms of 3D spatial information and time evolution. Real molecular clouds in spiral galaxies (\citealt{2007ApJ...668.1064E}) appear as high density regions in the ISM during the passage through the spiral arms. Works like \cite{2006MNRAS.371.1663D}, \cite{2012MNRAS.425.2157D}, \cite{2013MNRAS.430.1790B} were first attempts to set up more realistic initial conditions in simulations, which could have large effects on how clouds are shaped and how star clusters are forming.

There are large numbers of observations being made which contribute towards understanding the processes of stellar cluster formation. Observations are necessarily wide ranging, covering the mass-radius relation  (\citealt{2012A&A...543A...8M, 2016A&A...586A..68P, 2017A&A...601A..60C}), stellar age spreads (\citealt{2014ApJ...787..109G}, \citealt{2015ApJ...812..131K}), line of sight kinematics (\citealt{2008ApJ...676.1109F}, \citealt{2016A&A...589A..80H}), distribution of positions (\citealt{2008ApJ...686L.111K}, \citealt{2009ApJS..184...18G}), spatial structure (\citealt{2014ApJ...787..107K}, \citealt{2015ApJ...802...60K}, \citealt{2015ApJ...812..131K}). However, very few properties can be compared directly between simulations and observations. Simulations are limited by initial conditions and the physics included, while observations are limited by 2D projection in the sky as well as their inability to show us the the past or future evolution of clusters.

Star formation is also likely to be affected by feedback from the stars, especially young high mass stars. Feedback may also help to explain the low efficiencies of star formation seen on different scales (e.g. \citealt{2016AJ....151....5M, 2008A&A...477...79D, 2011ApJ...731...41O, 2016MNRAS.456.3432G, 2017MNRAS.466.1903G}), but low efficiencies are also possible in the absence of feedback (\citealt{2011MNRAS.410.2339B, 2014A&A...570A..15L}). Simulations of star formation including feedback generally result in a significant (factor 2) decrease in the star formation efficiencies (e.g. \citealt{2012MNRAS.422.1352D, 2014MNRAS.442..694D, 2015MNRAS.451..987D, 2015A&A...573A.112M, 2017MNRAS.467.1067D}). Although feedback can decrease the efficiency of star formation, it does not in general stop ongoing star formation. Instead, the feedback finds  weak points (lower density) in the surrounding gas through which it can be channelled and escape the dense clump.

Gas removal can affect the overall dynamics and lifetimes of forming clusters if it contributes a dominant proportion of the cluster mass. Several N-body studies have investigated the effect of gas removal in young stellar clusters via an assumed potential. These studies, although not fully consistent in terms of the effect of feedback on gas, provide valuable insight into the potential cluster evolution (\citealt{2001MNRAS.321..699K, 2012A&A...543A...8M, 2017A&A...600A..49B, 2017A&A...597A..28B}). For example, \cite{2015MNRAS.447..728B} showed that gas and dissipation free hierarchical mergers have difficulty producing large smooth clusters and that systems, such as R136, NGC3603 and the ONC could have formed from monolithic collapse scenario. However, in these works gas was replaced by static potential, neglecting the dissipational properties of gas dynamics which can greatly decrease the merger timescale (\citealt{2011MNRAS.410.2339B}). Indeed, several observational studies find evidence for hierarchical mergers in terms of kinematical subgroups in young stellar clusters (\citealt{2012ApJ...754L..37S}.).

R136, NGC3603 and the ONC cases does not necessarily confirm if these clusters has formed from monolithic collapse or internal structure, produced by mergers has been smoothed out by the interaction with gas. In contrast, observational studies find evidence of the highly fragmented nature of cluster formation (\citealt{2015A&A...581A.119B, 2017A&A...601A..60C, 2009ApJ...696..268Z}), for kinematical subgroups in young stellar clusters (\citealt{2012ApJ...754L..37S}) and of clusters that appear in close proximity such that a subsequent merger is possible, pointing to scenarios where hierarchical mergers are a likely process in star cluster formation (\citealt{2003MNRAS.343..413B, 2015MNRAS.449..715W}). These observational evidences brings support to cluster formation through merging scenario, which we will investigate in details in this work.

\begin{figure}
	\includegraphics[width=\columnwidth]{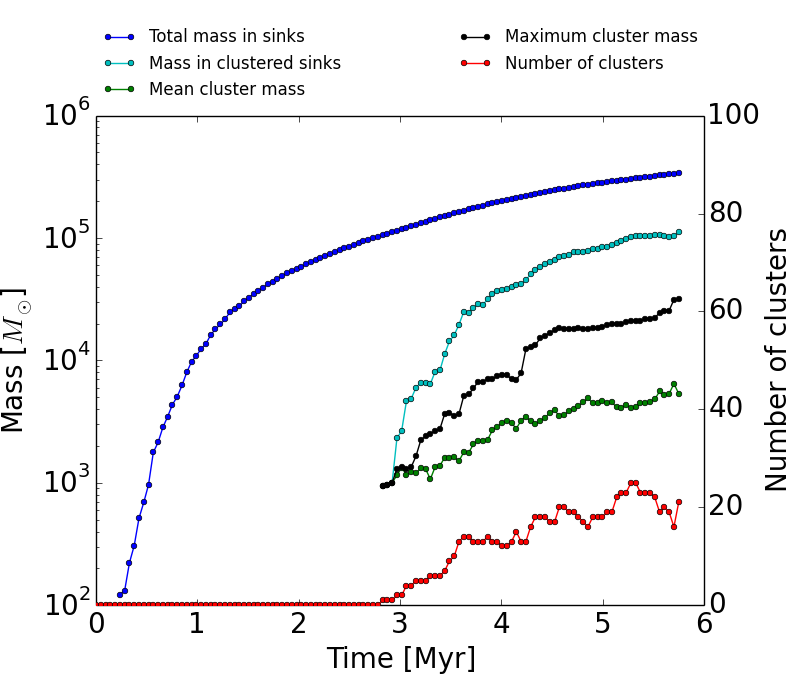}
	\caption{The sink and cluster statistics are plotted over the simulation. Star formation, modelled by sink particles, starts early (0.5 Myr), but clusters appear only from 3 Myr. The number of clusters is shown on right hand side axis. The mass in sinks and in clusters is also plotted (scale at left), showing continuous star formation throughout the simulation.}
	\label{fig:3o}
\end{figure}

\begin{figure*}
	\includegraphics[width=2\columnwidth]{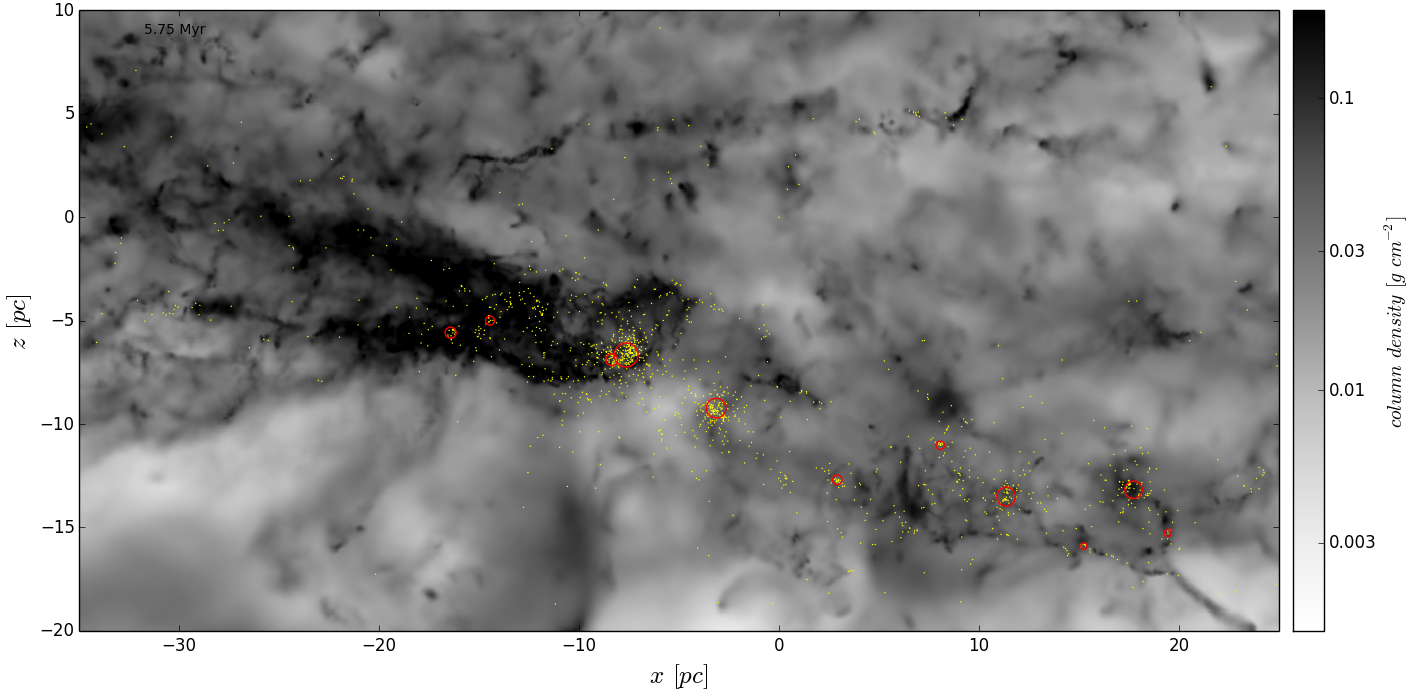}
	\caption{A side-on view through main star forming region in Galactic spiral arm is shown at at the end of the simulation. The LOS here is pointing through the Galactic plane, along y direction. The z direction is perpendicular to Galactic plane, where z=0 pc is the Galactic equator plane. The figure shows greyscale column density map for gas, the distribution of recently formed stars (in terms of sink particles) as yellow dots, and the positions of stellar clusters as red open circles. The map shows that dense gas clouds and forming clusters do not necessarily lie in Galactic mid-plane.}
	\label{fig:2}
\end{figure*}

\begin{figure*}
	\includegraphics[width=2\columnwidth]{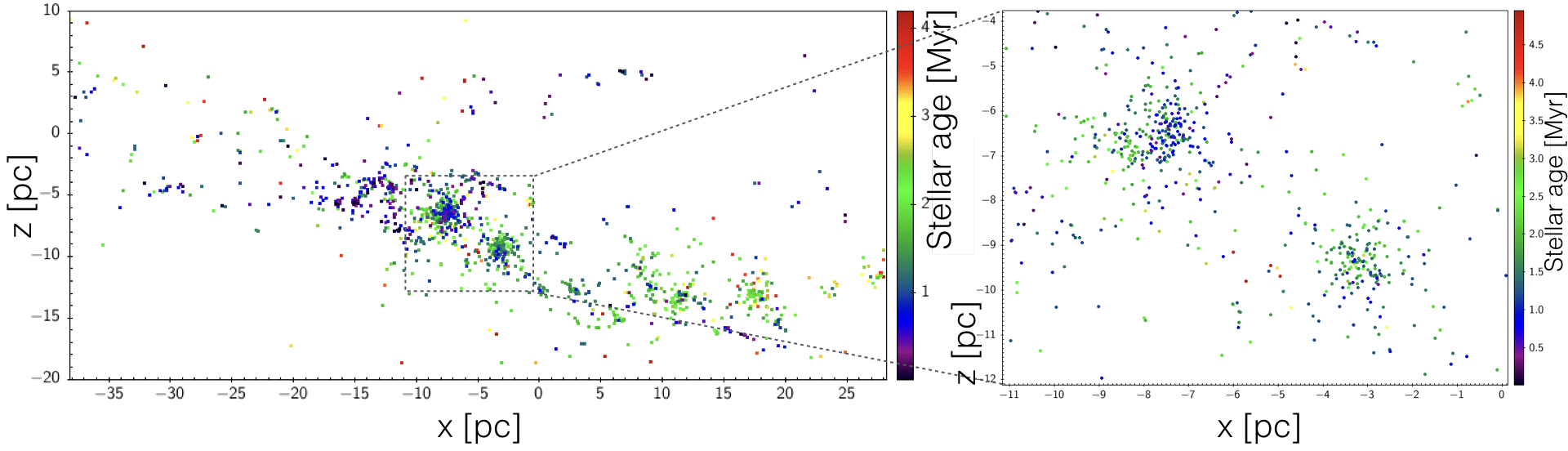}
	\caption{A side-on view through the main star forming region in the Galactic spiral arm is shown at at the end of simulation. The figure shows sink particles colour-coded by their stellar age. We see a clear age gradient from left to right which coincides with a decrease in the column density of gas in the same direction. This is evidence of a sequential star formation process triggered by the passage of the ISM through the spiral shock.}
	\label{fig:2s}
\end{figure*}

 \begin{figure*}
	\includegraphics[width=2\columnwidth]{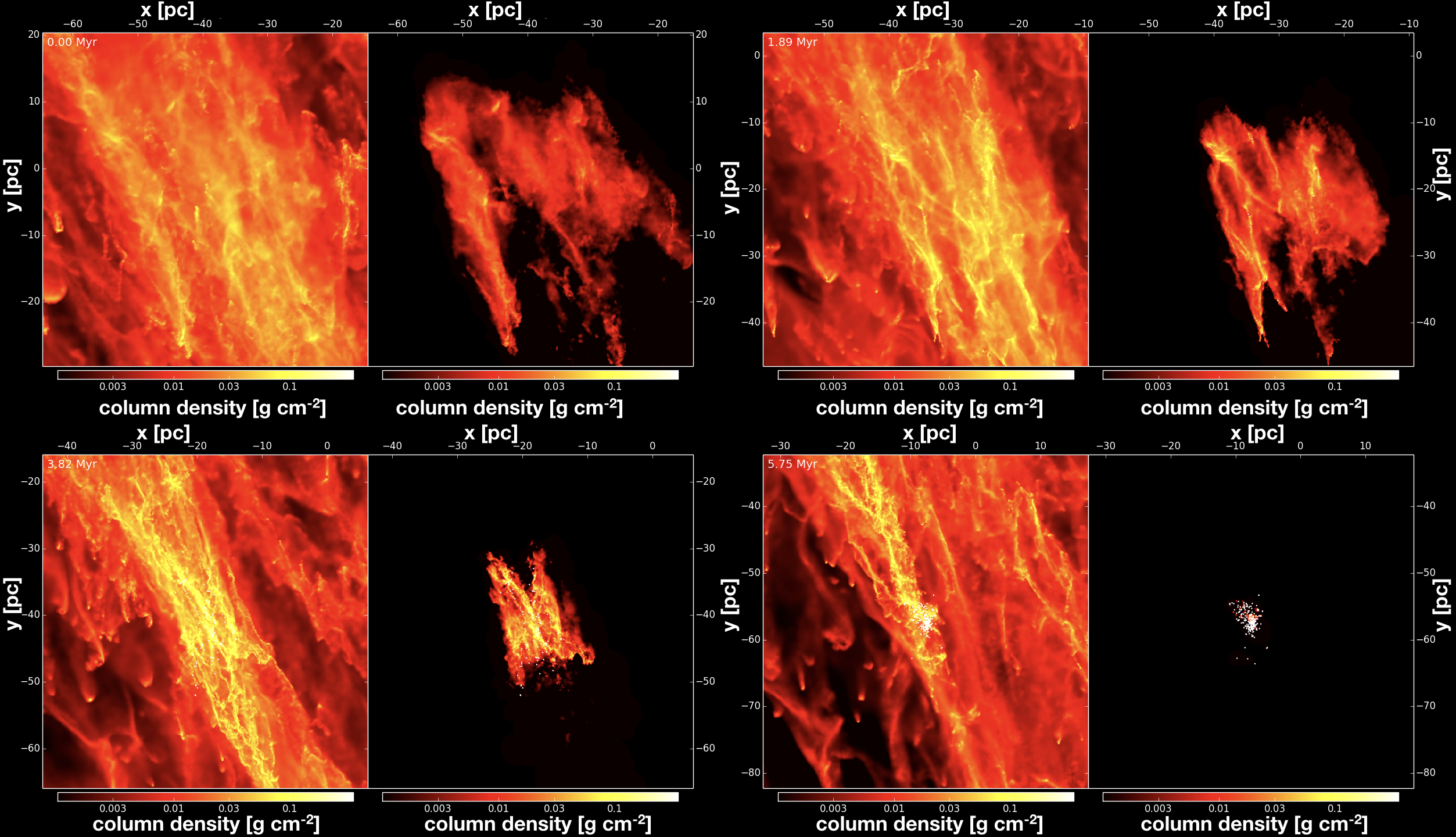}
	\caption{The face-on view evolution of the most massive cluster forming region calculated at four different times. The left hand panel in each pair shows column density of all gas in the region, while the right hand panel shows only gas accreted by the cluster. Maps show the compression of the ISM by the spiral shock passes. The central highest density region becomes gravitationally bound and forms a $\sim$30000 M$_{\odot}$ cluster.}
	\label{fig:clustEvol}
\end{figure*}

\section{Methods}

\subsection{SPH simulations}

In a previous work we analysed triggering of star formation during the spiral arm passage \cite{2016MNRAS.459.1985S}. Here we use the simulation data of \cite{2013MNRAS.430.1790B} which uses assumed spiral potential from \cite{2006MNRAS.371.1663D} through which ISM is allowed to flow. We determine physical cluster properties, such as masses, densities, sizes, binding energies and their evolution over time. These key parameters allow us to investigate stellar cluster formation mechanisms and dynamics free from the intrinsic assumptions due to the idealised initial conditions.

\cite{2013MNRAS.430.1790B} simulation was constructed from a set of nested simulations starting from a full Galactic disc simulation over 350 Myr, with a 50 Myr high-resolution counterpart focusing on the formation of dense clouds in the spiral arms,
and a final simulation to follow star formation over $5.8$ Myr. The final
stage, which we analyse here,  included self-gravity and modelled a 250 pc region containing $1.9\times 10^6$ $M_{\odot}$ mass.
This simulation used $1.29\times 10^7$ SPH particles with 0.15 $M_{\odot}$ masses. Star formation is followed through the use of sink particles (\citealt{1995MNRAS.277..362B}). A minimum mass for the sink particles corresponds to $\approx 70$ SPH particles representing one SPH kernel, or $\approx 11$ $M_{\odot}$ with a sink radius of $0.25$ pc to accrete bound, infalling gas particles while all particles penetrating within $0.1$pc were accreted. The sink particles therefore do not represent individual stars but rather a small cluster of stars or star forming region. Gravitational interactions between sinks were smoothed within $0.025$ pc.

Simulations used for our work, do not have any stellar feedback or magnetic fields, and is hence designed to investigate how stellar clusters are forming in more realistic initial conditions, which includes spiral arm dynamics. We also point out here that before including feedback and magnetic fields, it is essential to understand what properties of forming clusters are being defined purely by the spiral arm dynamics. Stellar feedback could be more relevant for later stages of simulation, once massive stars form, while at early stages only magnetic fields can slow down the formation of the first stars.

\subsection{Cluster definition}

In order to follow the early cluster evolution, we require a robust cluster definition and finding algorithm. There are a large number of cluster finding algorithms created to find clusters in datasets. For SPH simulations data, the cluster finding algorithm has to find clusters based on sink particle positions and masses. There are two steps to follow in order to obtain a robust cluster definition and to follow cluster formation history: static cluster definition at a given time and dynamical cluster definition by linking the same cluster from one time step to another.

Our cluster definition is based on sink local and enclosed gravitational potentials. Firstly each sink has assigned its local potential from surrounding neighbour sinks within 2 pc. The list of all sinks at a given time step is sorted by these potentials and the deepest potential sink is picked as a starting point. This first and lowest local potential sink is set as a centre of the cluster and enclosed potentials are calculated on all sinks within 2 pc around it. Next sinks are continuously added to the most bound cluster if their local potential is not deeper than twice that of the enclosed potential of the cluster.
If no such cluster is found and the sink has its local potential depth below the upper potential threshold ($-10^{11}$cm$^2$s$^{-2}$), it starts a new cluster. The search is finished when it reaches the first sink with its local potential above the background level. Clusters, which have fewer than six members are removed after the search is completed. Using the enclosed potential to build clusters results in a bias towards the selection of spherical rather than filamentary structures. We use $-10^{11}$cm$^2$s$^{-2}$ potential threshold, as systems below this potential becomes self gravitating and show virialized motions. We limit the cluster algorithm to not add any sinks whose local gravitational potentials are twice or less that of the enclosed potential of the cluster. Radial density profiles show that this ensures that we do not merge multiple clusters that are close to overlapping. In addition we use 0.05 pc softening for potentials in order to define smoother local potentials and remove unwanted fluctuations of potentials in cluster centres, as it can split one cluster into multiple if sharp peaks are detected.

We repeat the cluster finding process throughout the entire duration of the simulation at each time step. Clusters can be traced over time by linking two clusters between two neighbouring time steps. A cluster is assumed to be the same if particles representing more than 50\% of the cluster mass from the current time step $t_i$ are found in it at the next time step $t_{i+1}$. Clusters can also merge. If the mass of the smaller cluster is larger than 30\% of the total mass of both parts, then these clusters are major mergers (otherwise they are minor mergers). Mergers can be traced by searching if most of the sinks from two separate clusters at time $t_i$ are found in a single cluster at the next time step $t_{i+1}$. Linking clusters over all time steps allows us to create a merger tree for the clusters.

In order to remove fluctuations that occur when a cluster is near the cluster definition boundary, we smooth the cluster finding algorithm over neighbouring time steps. Cluster lifetimes are calculated along the merger tree branch. If the lifetime is only 1 time step, clusters are checked to ensure they do not merge again into the same cluster - if so, sinks from the temporary cluster are re-assigned to the main cluster. 

 \section{Results}
 
We make use of the \cite{2013MNRAS.430.1790B} self-gravity simulation that inherited its initial conditions from a large scale Galaxy simulation. It contains $\sim$ 12 million SPH particles. The inherited initial conditions set the galactic scale shock, and compresses the ISM to higher densities to start star formation. The gas motions into the shock are predominantly along y-axis of the simulation. Star formation takes place continuously as the shock travels through the gas, forming 2000 sinks and around 20 clusters by the end of the simulation. In addition, the shock reaches one side slightly earlier than another, and thus creates a gradient in stellar ages along the spiral arm. 
 
 \subsection{Cluster statistics}
 
We apply our cluster finding algorithm to the entire \cite{2013MNRAS.430.1790B} self-gravity simulation in order to find clusters of sinks. Clustering statistics based on the gravitational potential definition are shown in Figure \ref{fig:3o}. From the figure we see that at the beginning of the simulation there are no sinks and no clusters. The first sinks form quite early, in the first Myr of the simulation. However, clusters appears only around 3 Myr. This show that the first sinks form individually in the highest density fast collapsing clumps. Clusters appear slightly later when at least 6 sinks assemble in a compact region and meets the definition. Figure \ref{fig:3o} shows that star formation is continuously occurring in the simulation, with individual sinks, and clusters growing in mass, and in numbers throughout the simulation. The number of clusters appears to be growing up to $\sim$20 at the end of simulation. From $\sim$4 Myr fluctuations appear in the number of clusters, indicating that clusters also undergo mergers which add to their growth rates but decrease the total numbers of clusters present. 

\subsection{Cluster forming regions}

Figure \ref{fig:2} shows the large scale view of the gas, sinks and clusters at the end of simulation (5.6 Myr). The plot, viewed from the plane of the galaxy, shows many clusters lie in or nearby high density regions but do not necessarily match them. In Figure \ref{fig:2s} we plot the sink particles alone within the same limits, colour-coding them by mean stellar age. The plot shows a clear stellar age gradient. Sinks in the right hand side are slightly older because the spiral shock reached that side slightly earlier and triggered star formation there. As the spiral shock passed through, part of the gas was consumed by star formation, while another part moved with the shock or even started to expand as post-shock leaves the region. We can see very high column densities in the Figure \ref{fig:2} between $-30$ [pc] $< x < -10$ [pc], where the shock is moving through the gas. There is very little star formation in this part. However, looking at Figure \ref{fig:2s} we can see that star formation here only starts to happen with visible several very young (< 0.5 Myr) clusters (around $[-15; -5]$), which may merge in the future and form another big cluster there. Figures \ref{fig:2} and \ref{fig:2s} gives a support towards both sequential (\citealt{1977ApJ...214..725E}) and triggered star formation models: the sequential star formation appears as the spiral shock continuously propagates to the left but at the same time individual clouds are compressed, where star formation is triggered.

Most of the high density star forming regions in Figure \ref{fig:2} are visible 5-15 pc below the Galactic plane. This is due to the larger scale dynamics that drive the star formation (\citealt{2013MNRAS.430.1790B}), where spiral shocks and cloud-cloud collisions can cause some star forming regions to be significantly out of the plane of the Galaxy.

Right hand side panel of Figure \ref{fig:2s} shows zoomed in region onto two most massive final clusters. Cluster at left shows some younger stars in the core, but also a few older ones. Stars near the outside are mostly older. Cluster at right shows younger stars in the outer regions, some younger in core but overall a bit older.

We take the most massive cluster at the end of the simulation, which is visible on the edge of high density clouds in the middle of Figure \ref{fig:2}, and trace all its environment, accreted gas and sinks backwards in time. By finding sinks which belong to the cluster, we are also mapping all the gas particles that ultimately contributed to form the cluster. As all accreted particles are found, the cluster mass is assumed to be conserved over all time steps and the centre of mass of the system is well defined. We follow this cluster mass centre to illustrate the formation of the cluster in x-y position maps at four different time steps. Figure \ref{fig:clustEvol} shows the evolution of the gas and sinks forming the most massive cluster, plotted in the cluster's centre of mass frame. Each of 4 panels has a map of surface density derived from all gas particles in the region (left) and a map of surface density calculated from not yet accreted gas particles that contribute mass to the final cluster (right).

Initial gas cloud with the size of $\sim$40 pc is contracting down to several pc size cluster, a contraction of > 10 times over 6 Myr. Comparison between left and right hand side panels show that accreted cluster gas are embedded all the time in the largest density areas of the region. Accreted gas distribution (right panels) well matches the distribution of all gas (left panels). However, most of the particles visible in the left hand panel are not accreted by any sinks and so do not play a part in forming the cluster. Gas particles accreted by sinks not belonging to the cluster are excluded (right panels). The internal geometry of the region is highly fragmented with visible clumps (where the first sinks form) and filaments. Even if the cloud collapses as the whole, the internal structure of clumps and filaments keeps changing over the time. The presence of the galactic spiral shock is visible in the left hand panels, as the shock compresses widely distributed clouds to form a thin ridge through 6 Myr, extending from the top left to bottom right side of the map.

\begin{figure}
	\includegraphics[width=\columnwidth]{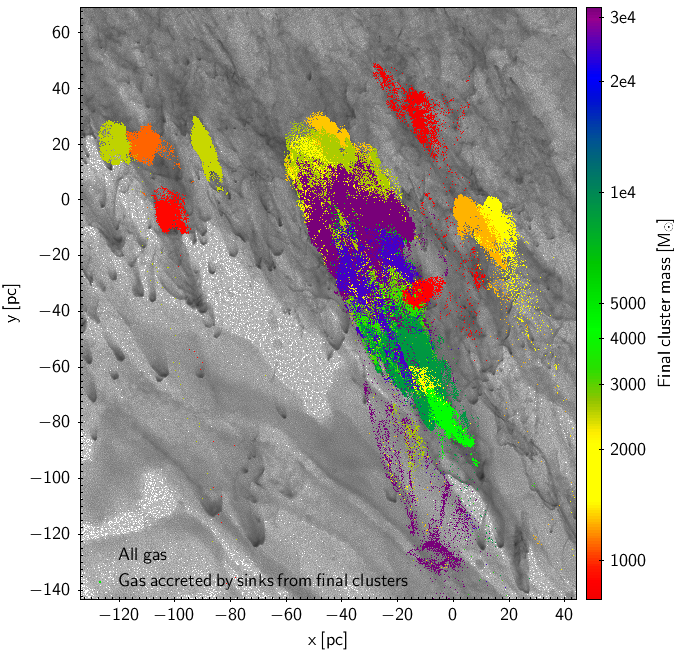}
	\caption{The face-on view of the main star forming region showing the initial size scales of the star forming clouds. Individual cluster forming reservoirs are colour coded by final cluster mass and plotted on the top of a greyscale map of all gas in the region. The map shows that clouds are aligned to the main ridge and that mass is coming into clusters from 10-30 pc size regions.}
	\label{fig:3}
\end{figure} 
 
Figure \ref{fig:3} shows the initial regions from which the various clusters form and accrete their final mass. These gas particles are shown in Figure \ref{fig:3} as coloured particles, with the colour representing the mass of the final cluster. The grey particles in Figure \ref{fig:3} show the non cluster-forming gas. We note that cluster forming gas clouds are initially embedded in higher density regions. The highest mass clusters form from the central high density region, and there are small mass clusters forming outside this region. We also notice the trend that more massive clusters form from larger clouds. The spiral shock is coming from the bottom-right side of the diagram and there are visible trails of particles, which are coming with the shock to the main star forming region and are also accreted by sinks in clusters (purple trails in the bottom side of the diagram). As the material is incoming with the small pitch angle to the spiral arm, the section of low density inter-arm gas can be seen in the bottom-left side of the diagram.
 
\begin{figure}
	\includegraphics[width=\columnwidth]{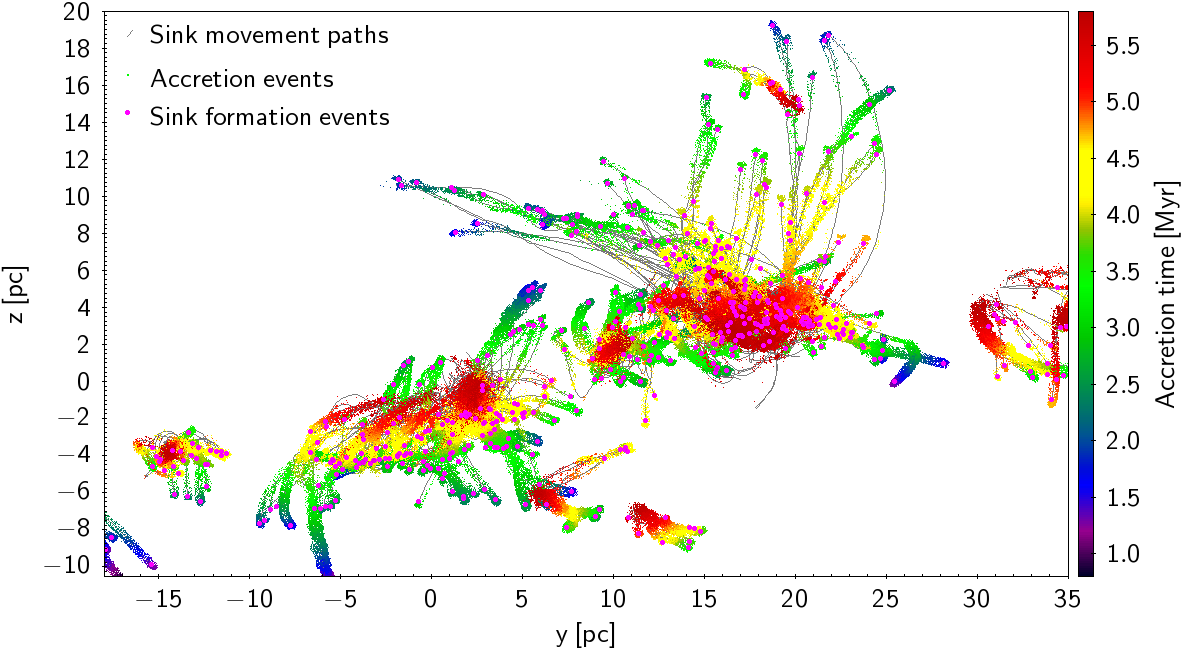}
	\caption{This diagram shows the location of star formation and subsequent accretion of gas particles during the cluster formation process.}
	\label{fig:5}
\end{figure}

\subsection{Accretion histories}

The Lagrangian nature of the simulation allows us to trace and reconstruct accretion maps in an unprecedented level of detail. The accretion histories for the two most massive clusters are shown in Figure
 \ref{fig:5}. Most massive cluster is visible slightly to the right from the centre of the picture, while the second most massive cluster - slightly left and down. Particle positions are plotted relative to the centre of mass frame of all clusters in the simulation. Accreted gas particles are plotted at their final locations, and colour-coded  with the time of their accretion. Sink forming locations are plotted as purple dots. Sink movement paths are shown as grey lines.

This figure shows the formation process of the two clusters. The first sinks are forming in relatively isolated regions. They form along filaments, and then flow down the filaments. Additional sink formation forms small clusters which grow through accretion, and mergers. Star formation continues along the filament and down to the intersection points where the filaments flows, and accompanying clusters merge to form the final cluster. 
 
\section{Cluster properties}

In the following sections, we analyse the developing properties of the stellar clusters in our simulation. It should be noted that as the simulations do not include feedback, the properties could be affected (\citealt{2013MNRAS.432..986P, 2015MNRAS.447..728B}). 
\cite{2013MNRAS.432..986P} uses \cite{2012MNRAS.424..377D, 2013MNRAS.430..234D} simulations to show that clusters formed with feedback have only slighlty smaller masses, bigger sizes and larger dynamical times.
 
\subsection{Mass merger tree}

\begin{figure*}
	\includegraphics[width=2\columnwidth]{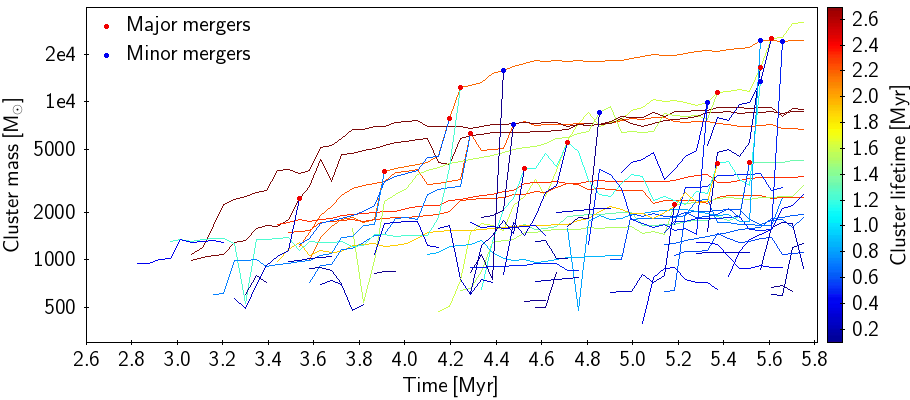}
	\caption{The diagram showing the mass-merger tree of clusters. The diagram shows major (smaller cluster has more than 30\% of total cluster mass after merging) and minor merging events. More massive clusters also have longer lifetimes and more extended merging histories.}
	\label{fig:6}
\end{figure*} 

We trace clusters between multiple time steps in order to follow the evolution of cluster properties over their formation histories. The simplest cluster property is its mass. As we know which sinks are members of which cluster, we obtain cluster masses over multiple time steps. Multiple cluster events can occur over time, such as creation, dissolution, merging and splitting. As our simulation is targeting cluster formation process, we naturally have merging processes of two clusters occurring at particular time steps. Plotting the cluster masses over time produces a mass-merger tree diagram (Figure \ref{fig:6}). The intersection points show major merger events as red, if the child cluster's mass exceeds 30\% of the parent's cluster mass. Minor merger events occur when the child's mass is below 30\% of the parent cluster's mass and are plotted as blue points. Colours show the total lifetime of the cluster, from its formation until it merges with the parent cluster.

Firstly we notice that cluster mass growth and merging continues throughout the simulation. Clusters continue to grow as long as there is surrounding gas, sinks or smaller clusters to be accreted into larger central clusters. Merger events occur most frequently for larger mass clusters, while low mass systems have small numbers of mergers in their histories. Cluster merging also appears to be a channel of significant mass growth over the time for the more massive systems. Large final mass clusters have also much longer merging histories, well traced over $\sim$2.5 Myr, while small clusters have lifetimes of only 0.5 - 1 Myr.

\subsection{Mass-radius relation}

\begin{figure}
	\includegraphics[width=\columnwidth]{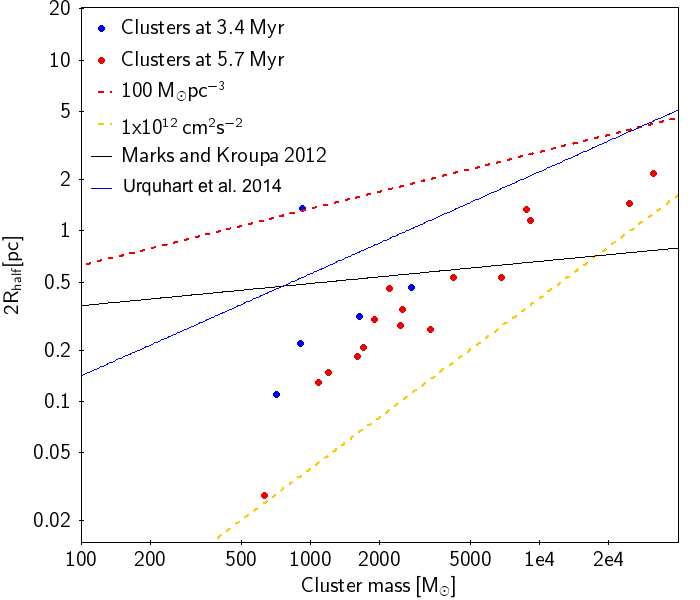}
	\caption{The cluster mass-radius relation is shown at two different times of the simulation. The plots shows how the size of the clusters increases with the mass of the cluster, with an approximate R$_{half}$ $\sim$ M$_{clust}^\gamma$ relation, where $\gamma$ is about 1.}
	\label{fig:7}
\end{figure}

One of the properties of our simulated clusters is their mass-radius relation. This must reflect the formation process in some way and thus should form a good property with which to compare to real clusters. As cluster definition returns a list of cluster members, the characteristic cluster sizes can be determined. We use twice the half-mass radius in order to characterise full cluster sizes. The half-mass radius for each cluster is found by sorting cluster member sinks by their distances from the cluster's centre of mass. We then use cumulative enclosed mass radial profiles to find at what radius half of the cluster mass is enclosed.  We plot twice the half-mass radius as a robust measure of the effective size of the cluster that does not suffer from the variation in position or classification of the outermost cluster members. When cluster masses and sizes are determined we get a cluster mass-radius relation. We then plot cluster mass-radius relation over all clusters at two time steps - blue for early times (5 clusters) and red for late times (16 clusters) (Figure \ref{fig:7}).

The diagram shows that more massive clusters also have larger half-mass radii. Low mass clusters, taken to be those with $<$1000 M$_{\odot}$ have half-mass radii of 0.1 pc. On the other hand high mass clusters have half-mass radii of 0.5 - 2 pc. We found clusters between the -10$^{11}$ cm$^2$s$^{-2}$ iso-potential line and 100 M$_{\odot}$ pc$^{-3}$ iso-density line.
On one hand this looks as artificial disadvantage of the definition, as clusters are more likely to be traced as spherical systems. But on the other hand these spherical and centrally condensed systems are virialized and do not change their geometrical shape rapidly in time.

Several physical processes are important for the mass-radius relation. These include the gravitational collapse that forms the cluster, subsequent merger events and ongoing gas and sink accretion. Gravitational collapse causes the size of the system to decrease. On the other hand, mergers grow the cluster mass but there is also an increase in the combined cluster size.

Blue solid line shows mass radius relation from fitting observational data of star forming clumps by \cite{2014MNRAS.443.1555U}. The line appears to be slightly above our data points. This could be because \cite{2014MNRAS.443.1555U} mass-radius relation was fitted for clumps, which are still collapsing. If clumps would continue to collapse towards clusters, their radii would decrease and could match our simulated clusters.

We also plot a black solid line in Figure \ref{fig:7} which shows \cite{2012A&A...543A...8M} theoretically predicted birth mass-radius relation, obtained by using binary populations in clusters. \cite{2012A&A...543A...8M} mass-radius relation is derived for the densest collapse state of the bulk young stellar population in the cluster, and how these limit the binary statistics. This indirect measurement of the mass-radius relationship is powerful but inherently makes assumptions as to the natal binary properties. Subsequent dynamics and tidal evolution will likely affect the cluster properties (\citealt{2012MNRAS.425..450M}).

\subsection{Cluster angular momentum}

\begin{figure}
	\includegraphics[width=\columnwidth]{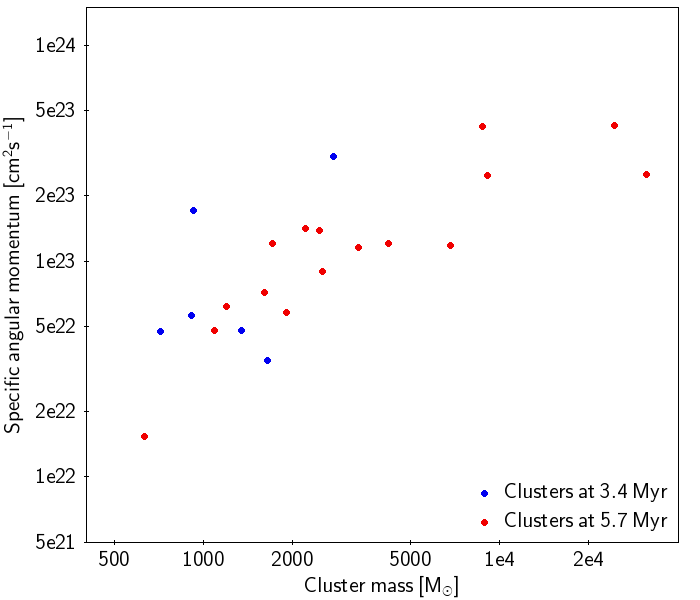}
	\caption{The stellar clusters' specific angular momentum is plotted as a function of cluster mass. Higher mass clusters have larger specific angular momenta, indicating that rotation, due to accreting mass from further away, is more important in such clusters.}
	\label{fig:9ang}
\end{figure}

Due to the self consistent initial conditions used in this study, we can make a first estimate of the angular momentum of the newly formed clusters. Figure \ref{fig:9ang} shows specific angular momenta of these clusters as a function of their masses at early (blue points) and late (red points) time steps. We use all members of the cluster, relative to the centre of mass of the system when calculating angular momenta.

We note that larger mass clusters also have larger specific angular momenta. This could be a result of collecting gas and hence angular momentum from larger scales. Merging processes in clusters involve contributions from larger scales. Merging two clusters results in a jump in this diagram towards larger masses and larger specific angular momentum. Even if Figure \ref{fig:9ang} shows higher specific angular momentum, the rotational contribution is only several to several ten percent.

\subsection{Mass growth of stellar clusters}

\begin{figure}
	\includegraphics[width=\columnwidth]{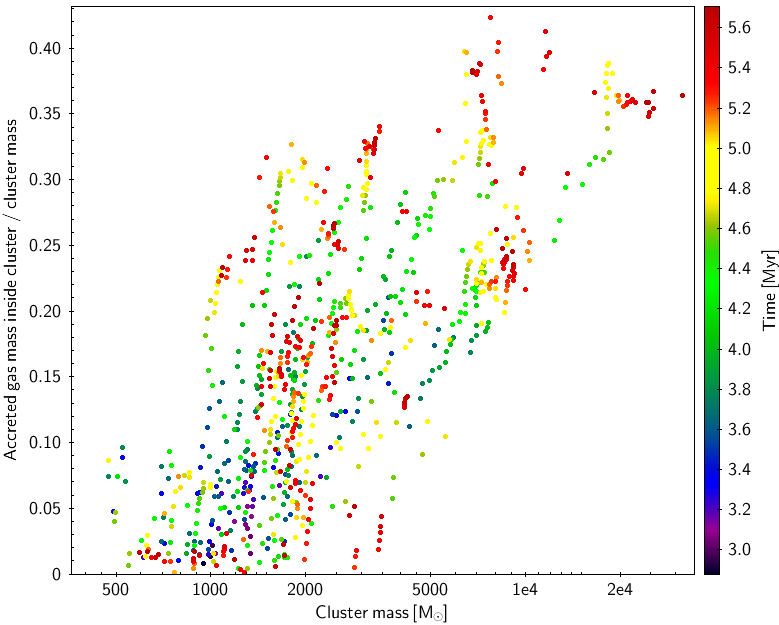}
	\caption{The growth in clusters' mass due to accretion of gas inside the cluster. Gas accretion is an increasingly important contributor to cluster masses as a function of mass. This is due to the larger range from which they can accrete due to their deeper gravitational potential.}
	\label{fig:9}
\end{figure}

Accretion of stars (\citealt{2007MNRAS.375..855P}) and gas (\citealt{2009MNRAS.397..488P}) from the environment has been investigated in pre-existing clusters. Here we address what contributes mostly in forming different mass clusters - accretion of stars or gas. In order to measure the accretion of gas into the cluster, we follow the change of mass of the sinks that are already in the cluster. In Figure \ref{fig:9}, we plot this as a fraction of the cluster mass and as a function of the total cluster mass.
 
The first thing we notice is that clusters with low masses appear to have only a small fraction of their mass being due to gas accretion. This is partially by definition in that the cluster first "forms" when six or more sinks are sufficiently close, and hence have a minimum mass to which gas accretion does not contribute. Nevertheless, we see that the fraction of the total mass contributed by gas accretion increases with cluster mass. At high cluster masses, direct accretion of gas is seen to contribute nearly half the total mass of the system. 
Secondly, we notice that mass gain from gas accretion inside clusters is always less than 40 - 50 \% of the total cluster mass (This neglects the gas accretion onto the initial cluster formation consisting of a minimum of several hundred solar masses). The diagram clearly shows that gas accretion on clustered sinks has contributed mostly for large mass clusters.

If the cluster is not accreting any sinks but only grows by gas accretion on its existing sinks, then it moves upwards to larger accreted gas mass over cluster mass ratios and also larger cluster masses. This is visible as a forest of parallel trails going upwards in Figure \ref{fig:9} for low mass clusters at 2-4 Myr. If cluster accretes sinks or other clusters, they "jump" downwards, which is visible for larger clusters at later times (4-6 Myr).

\subsection{Cluster age spreads}

 \begin{figure}
	\includegraphics[width=\columnwidth]{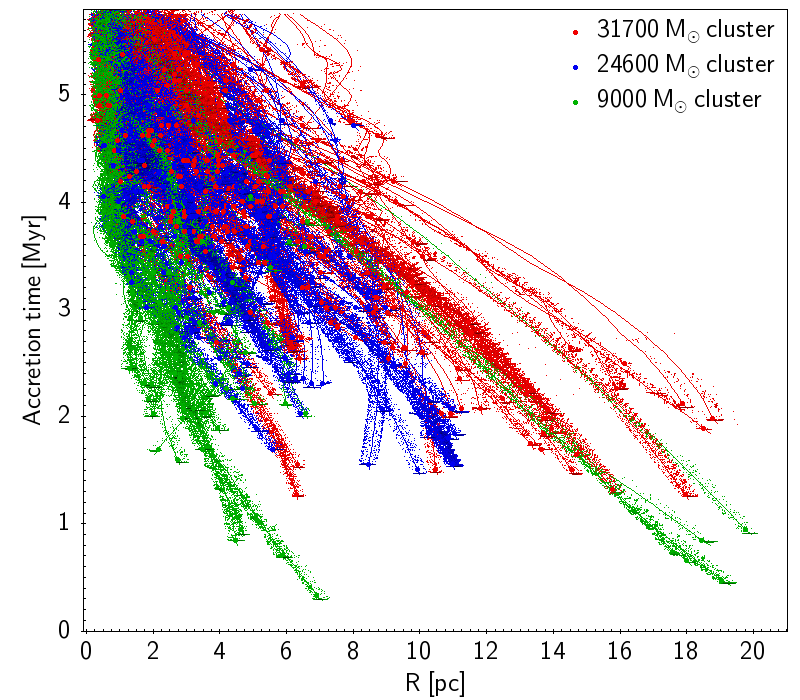}
	\caption{The accretion timeline is shown for the three most massive clusters. Sink formation and accretion extend over several Myr as the clusters form from star formation that extends more than 10 pc from the centre of mass. Some of 9000 M$_{\odot}$ cluster sinks form as far as 18 - 20 pc away at very early 0.5 Myr simulation time, and comes into the cluster later.}
	\label{fig:10}
\end{figure}

 \begin{figure}
	\includegraphics[width=\columnwidth]{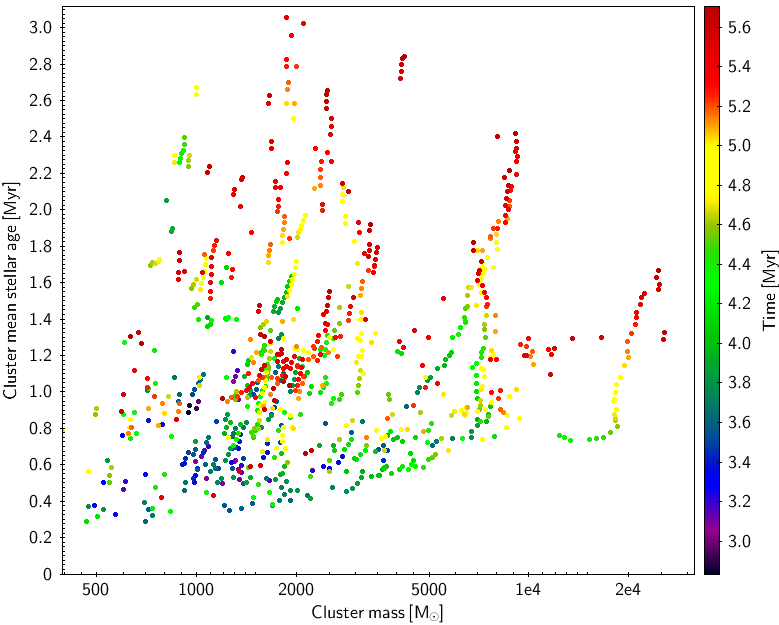}
	\caption{Clusters' mean stellar ages are plotted as functions of cluster mass and time in the simulation. The Mean ages change only slowly as long as star formation and accretion are ongoing. Once star formation ceases, the clusters appear to age normally, at near constant mass. The only mass growth is then limited to the merger of other clusters.}
	\label{fig:12}
\end{figure}

\begin{figure}
	\includegraphics[width=\columnwidth]{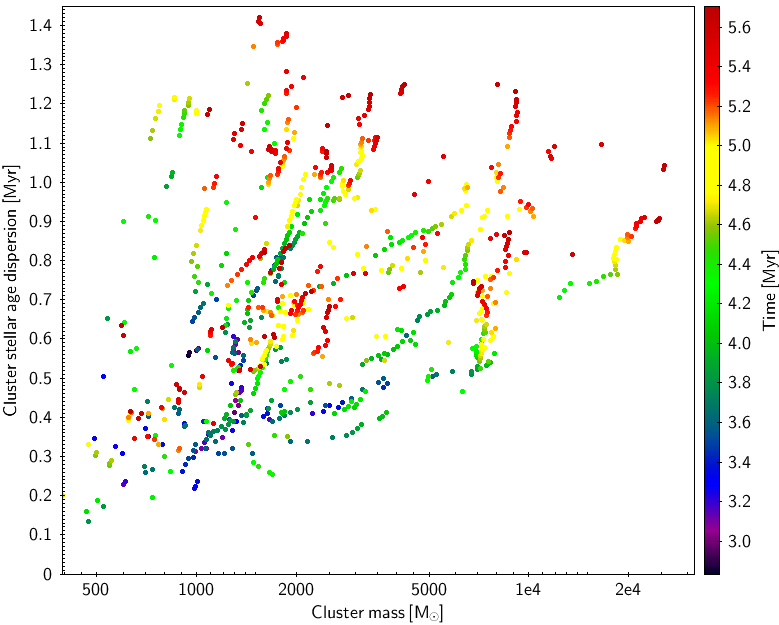}
	\caption{Cluster stellar age dispersion is plotted as a function of cluster mass and time. Low mass young clusters show low stellar age dispersion while large mass clusters have rich merging histories, resulting in larger stellar age dispersion.}
	\label{fig:13}
\end{figure}

The merging process leads not only to a growth in mass, but also to a naturally larger age spread as the resultant clusters are formed by mixing different systems formed in different environments. Figure \ref{fig:10} shows the accretion histories, in terms of accretion time versus the radius at which particle has been accreted. Here we plot only three most massive final clusters and trace their sinks backwards in time to where they formed. The common centre of mass for the system is calculated from all particles which make up the final mass of the cluster.

The sinks can be seen to consistently move to small radii, indicating continuous cluster collapse. We see that the cluster collapse is continuous over all time steps as sinks are continuously moving towards smaller radii. We also see that accretion takes place over several Myrs, in the absence of feedback. While feedback is often assumed to halt ongoing star formation, simulations show that star formation can continue due to the channelling of feedback away from the dense star forming gas (\citealt{2012MNRAS.422.1352D, 2014MNRAS.442..694D, 2015MNRAS.451..987D, 2017MNRAS.467.1067D}). The large dispersion in accretion times of various sinks originating in different regions produces significant spreads in stellar age in the final clusters. Smaller sub-clusters, which are visible as groups of paths, have smaller age dispersions. On the other hand, smaller clusters, which have yet to merge, display smaller age dispersions.
 
In Figure \ref{fig:12} we plot mean stellar ages for each cluster as a function of cluster mass and time. Stellar ages for each sink are calculated as a mass weighted average of accretion times for already accreted gas particles. This gives, for each sink, its stellar age, which can be different from the time since the sink first formed. The stellar age is smaller than sink age due to more recent accretion events. As we do not fully resolve star formation, we cannot distinguish whether this ongoing accretion could be forming additional stars, or accreting into pre-existing stars. Ongoing accretion onto young stars can also significantly reduce their apparent ages (\citealt{1999MNRAS.310..360T}). We noticed that in our simulation stellar ages are on average 2/3 that of sink ages. Finally the cluster mean stellar age is just an average of stellar ages for all its members. Figure \ref{fig:12} show the visible trails of individual clusters over different times. At early times, sinks in clusters are accreting intensively and thus, their mean stellar ages increase slowly.

Once accretion in the cluster stops (as occurs when most material is accreted), the mean stellar age starts to increase rapidly and cluster mass growth slows down. During this phase, the tracks can be seen to be almost vertical in Figure \ref{fig:12}. There is also a visible trend with mean stellar ages for early times slightly increasing for higher mass clusters. Stellar age dispersion similarly increases for more massive clusters, as they have experienced more mergers, bringing together stars which have formed at different times in a wider variety of environments.

We also show stellar age dispersions in Figure \ref{fig:13}, which are calculated as an age dispersion of all accretion events over all cluster members. The diagram shows a clear trend that stellar age dispersion increases for higher mass clusters. This is in a good agreement with merging scenario, as more massive clusters have experienced more mergers, bringing together stars of different ages from a wider variety of environments.

\section{Conclusions}

We performed an in depth analysis of how stellar clusters form in large Galactic scale simulations that resolve individual cluster forming regions, but neglect feedback processes. Galactic spiral shocks assemble clouds in the ridge-like structures, where we measured a stellar age gradient as the shock approaches one side of the ridge earlier than another. Older clusters are found in the regions that entered the spiral shock earlier. Younger clusters, and ongoing star formation, are associated with regions that more recently entered the spiral shock and hence are also associated with dense gas clouds.

Our analysis relies on a physically based cluster definition, which uses local and enclosed gravitational potentials in order to separate individual clusters. We noticed that a robust physical cluster definition can be one of the most vital steps before determining further physical properties and relations for stellar clusters.

The Lagrangian nature of SPH allowed us to trace individual cluster formation and accretion histories over time. We reconstructed accretion maps, showing details of how individual star forming clumps are moving in global gravitational potential, including where and when star formation is occurring. Clusters appear to form in separate local regions, which undergo collapse and experience frequent merging process over the time. We produced a cluster mass merger tree diagram which show that merging is an important process in massive cluster formation. Merging also produces stellar age spreads up to 1 Myr as material comes from different environments.

We include predicted cluster mass-radius relation showing how higher mass clusters are expected to be substantially larger. Our smallest resolved 1000 M$_{\odot}$ clusters show their half-mass radii of 0.1 - 0.2 pc while 20000 M$_{\odot}$ clusters have 1 - 2 pc half-mass radii. In order to resolve smaller mass clusters, higher resolution simulations would be needed. Analysis of angular momenta show that more massive clusters have higher specific angular momentum, which can be attributed to having to accrete from significantly larger volumes and hence higher velocity dispersions. We also address what drives cluster mass growth of different mass clusters. Less massive clusters appear to be growing by assembling locally formed sinks, while more massive clusters have powerful global gravitational potentials, which allow surrounding gas to be efficiently channelled to the cluster centres, accelerating accretion.

\section*{Acknowledgements}

RS and IAB acknowledges funding from the European Research Council for the FP7 ERC advanced grant project ECOGAL.
This work used the DiRAC Complexity system, operated by the University of Leicester IT Services, which forms part of the STFC DiRAC HPC Facility (www.dirac.ac.uk). This equipment is funded by BIS National E-Infrastructure capital grant ST/K000373/1 and  STFC DiRAC Operations grant ST/K0003259/1. DiRAC is part of the National E-Infrastructure.

We thank William Lucas, Felipe Gerardo Ramon Fox, Duncan Forgan, Rowan Smith and Claudia Cyganowski for helpful discussions and comments.




\bibliographystyle{mnras}
\bibliography{bibliography} 







\bsp	
\label{lastpage}
\end{document}